\documentclass[%
 reprint,
superscriptaddress,
showpacs,
showkeys,
amsmath,amssymb,
aip,
]{revtex4-2}
\usepackage{graphicx}
\usepackage{dcolumn}
\usepackage{bm}
\usepackage[breaklinks]{hyperref}
\hypersetup{
  colorlinks,
  citecolor=blue,
  linkcolor=black,
  urlcolor=blue}
\usepackage[outdir=./PS/]{epstopdf}
\graphicspath{{./PS/}}
\usepackage{tikz}
\usepackage{overpic}
\usepackage{color}
\usepackage{comment}
\usepackage{upgreek}
\usepackage[T1]{fontenc}
\usepackage[utf8]{inputenc}
\usepackage{soul,xcolor}
\setstcolor{red}

\DeclareMathOperator{\sgn}{sgn}

\newcommand{\vct}[1]{{\bf #1}}

\newcommand{\tu}[1]{\mathrm{#1}}

\newcommand{\Pc}{\mathcal{P}}
\newcommand{\Sc}{\mathcal{S}}

\newcommand{\Tc}{\mathcal{T}}

\begin{document}


\title{
  Surface polarization strongly influences electrostatics in a
  nonlocal medium}


\author{Ali Behjatian}

\affiliation{Physical and Theoretical Chemistry Laboratory, Department
  of Chemistry, University of Oxford, South Parks Road, Oxford OX1
  3QZ, UK.}

\author{Ralf Blossey}

\affiliation {Universit\'e de Lille, Unit\'e de Glycobiologie
  Structurale et Fonctionnelle (UGSF), CNRS UMR8576, 59000 Lille,
  France}

\author{Madhavi Krishnan}%
\email{madhavi.krishnan@chem.ox.ac.uk}

\affiliation{Physical and Theoretical Chemistry Laboratory, Department
  of Chemistry, University of Oxford, South Parks Road, Oxford OX1
  3QZ, UK.}

\affiliation{The Kavli Institute for Nanoscience Discovery,
  Sherrington Road, Oxford OX1 3QU, UK. }%


\begin{abstract}
  Electrostatics in the solution phase is governed by free electrical
  charges such as ions, as well as by bound charges that arise when a
  polarizable medium responds to an applied field. In a local medium,
  described by a constant dielectric permittivity, the sign of the
  far-field electrostatic potential distribution around an object is
  governed by its electrical charge. We demonstrate significant
  departures from this expectation in a nonlocal medium characterized
  by a wave vector-dependent dielectric function. Here, surface
  polarization due to the solvent, or indeed non-solvent dipoles, may
  wield significant influence at large distances. The polarization
  correlation length may not only significantly augment the effective
  screening length, but we also show that the electrical contribution
  from polarization can compete with and even invert the sign of the
  electrical potential and the field arising from charge alone. These
  results hold ramifications for a range of apparently anomalous
  electrically governed observations, such as underscreening,
  electrophoretic mobilities of charge-neutral objects, and
  long-ranged attraction between like-charged entities in water and
  other solvents.
\end{abstract}
                        
\maketitle

\section{Introduction}
Interactions between charge-carrying molecules and particles in the
fluid phase form the bedrock of collective processes underpinning
chemistry and biology. In media of high dielectric constant such as
water, and in the presence of low concentrations of monovalent salt
ions, the Poisson-Boltzmann (PB) theory is believed to provide an
accurate description of electrostatic interactions. A hallmark of
PB-theory is the expectation of a repulsive force between electrically
like-charged entities that decays with increasing separation at a rate
given by the Debye screening length,
$\kappa^{-1}= 1/\sqrt{8\pi n_0 \ell_\tu{B}}$. Here, $n_0$ is the bulk
number density of positive or negative monovalent ions in solution,
and $\ell_\tu{B} = e^2/4 \pi \epsilon_0 \epsilon_s k_\tu{B} T$ is the
Bjerrum length, where $e$, $\epsilon_0$, $\epsilon_\tu{s}$,
$k_\tu{B}$, and $T$, denote the elementary charge, permittivity of
free space, static dielectric constant of the medium, Boltzmann's
constant, and absolute temperature, respectively. Many important
indications of PB theory have, indeed, been captured in a range of
mechanical force measurement experiments
\cite{horn1989,ducker1992,trefalt2017}. However, there have been
wide-ranging experimental reports of departures from PB theory, mainly
centered on interactions inferred from the spatial structure of
suspensions of particles and molecules probed using visible light and
x rays \cite{klug1959,matsuoka1991,kepler1994,
  han2003,baksh2004,haro2009,wang2024}.  Furthermore, interesting
observations persist concerning the electrokinetic mobility of neutral
lipid vesicles, as well as long-ranged attraction measured for
net-neutral lipid bilayer coatings \cite{egawa1999,silbert2012}.

The canonical theoretical view of electrostatic interactions entails
treating a fluid as a local continuum described by a phenomenological
electric susceptibility constant, $\chi \ge 0$, with the solvent
merely providing a structureless shielding background
\cite{kornyshev1978}. The susceptibility constant, $\chi$, establishes
a linear relationship between the polarization, $\vct{P}$, and the
electric field, $\vct{E}$, i.e.,  $\vct{P}=\epsilon_0 \chi
\vct{E}$. Far from a structureless continuum, however, fluid media are
in fact grainy. The structure is sustained by short-range
intermolecular interactions such as dipole-dipole interactions and
hydrogen bonding, but may also display long-range orientational
correlations according to several experimental reports
\cite{shelton2014,chen2016,duboisset2018,dedic2019,dedic2021}.  Indeed
the incorporation of a description of intermolecular interactions led
to successful modeling of the short-range hydration force, which was
first experimentally observed in the 1980s
\cite{israelachvili1983,marcelja1976}. Here we construct a model of
electrostatics in a medium capable of sustaining molecular
orientational correlations that decay over a distance $\xi$ -- the
polarization correlation length. We examine the consequences thereof
on the electrostatic far-field, defined by distances $2$--$3$ times
larger than the Debye length, the relevant length scale in a local
medium or electrolyte.

A significant simplification at the heart of local continuum
electrostatic theories posits that the response of the medium,
$\vct{P}$, may not depart from that dictated by the disturbance,
$\vct{E}$.  Accordingly, the polarization surface charge, defined as
$\sigma_\tu{p} = \vct{P} \cdot \vct{n}$, where $\vct{n}$ is the normal
vector directed outward to the dielectric material
(Fig. \ref{fig:schm}), is determined self-consistently from Gauss' law
and cannot be treated as an independent parameter of the system.
Molecular dynamics (MD) simulations however reveal the presence of
oriented solvent dipoles in the vicinity of neutral
surfaces/interfaces immersed in a variety of polar solvents
\cite{reif2016,loche2018,walker2022,kubincova2020}. This occurs
presumably due to anisotropic bonding interactions of solvent
molecules located at a discontinuity and can entail substantial
magnitudes of net polarization, $P\approx0.1e$ nm$^{-2}$, even at a
neutral surface carrying no electrical charge, immersed, e.g., in
water. Such a behavior has also been extensively confirmed in nonlinear
spectroscopy experiments on water at interfaces
\cite{du1994,ye2001,shen2006,myalitsin2016}. This net orientation of
solvent molecules at interfaces thus implies an ``excess'' interfacial
polarization which has been suggested to play an important role in
interparticle interactions, but the effects of which cannot be
completely and self-consistently captured within the local
electrostatic view \cite{kubincova2020}.

In contrast, a nonlocal medium permits the polarization $\vct{P}$ to
depart from $\vct{E}$ over distances where polarization correlations
are non-negligible. A discontinuity in the polarization across the
interface, arising e.g., from anisotropy in solvent bonding
interactions, produces a polarization surface charge, $\sigma_\tu{p}$.
In an ion-free nonlocal medium, the polarization surface charge
couples into that in the bulk. This generates a spatial polarization
profile and concomitant electric field whose decay is governed by the
polarization screening length $\lambda \propto\xi$.  Thus for an
object immersed in an electrolyte, even in the absence of charge on
the object, an electrical double layer may form. When
$\lambda\gtrsim\kappa^{-1}$, the effective extent of the resulting
counterion cloud can depart significantly from that given by
$\kappa^{-1}$ characteristic of a local medium. Furthermore, in the
presence of both surface charge density, $\sigma$, and polarization
charge density $\sigma_\tu{p}$, coupling of the two sources results in
an overall net electric field whose sign, magnitude, and spatial rate
of decay are sensitively governed by the values of the underpinning
system parameters, as discussed in detail in what follows.
Importantly, $\lambda$ and $\xi$ are in general not identical, and the
relationship between these quantities strongly depends on the form of
the dielectric function which describes the nonlocal response of the
medium. While the determination and implementation of the exact form
of dielectric function are non-trivial tasks
\cite{bopp1996,bopp1998,monet2021}, the use of a Lorentzian
approximation can lead to a significant simplification of the model
\cite{kornyshevChemSol1985} and a linear relationship between
$\lambda$ and $\xi$: i.e., $\lambda = \xi \sqrt{\theta}$. Here,
$\theta = \epsilon_\tu{s}/\epsilon_\infty$ is the ratio of the static
dielectric constant of water, $\epsilon_\tu{s} \approx 80$, to a value
$\epsilon_{\infty} \approx 2$--$5$ at frequencies just above the first
dispersion beyond which the rotational modes are effectively frozen.

In the present study, we describe the consequences of this view based
on a straight-forward generalization of PB-theory.  We consider the
impact of interfacial solvent polarization and embedded interfacial
dipoles on the electrostatic far-field of an isolated charged plane in
contact with a nonlocal electrolyte. The implications of an excess
interfacial polarization for far-field electrostatics in nonlocal
media have to our knowledge not been extensively explored, especially
in the nonlinear regime where the Debye-H\"uckel (DH) approximation
breaks down.  We examine the results of the two-field formulation of
nonlocal electrostatics proposed in
Refs. \onlinecite{hildebrandt2004,paillusson2010} for surfaces carrying
electrical charge $\sigma$ in combination with (1) a surface
polarization charge, $\sigma_\tu{p}$, and (2) a dipole layer of
surface density, $\tau$, embedded in the interfacial region at a
distance $\ell$ from the surface (see Fig. \ref{fig:schm}).  The
latter case may be taken to describe zwitterionic lipid headgroups,
typically encountered in lipid monolayers or bilayers, or zwitterionic
surface coatings. Here we only consider the role played by charges
that describe the embedded dipoles and do not attempt to explicitly
capture solvent polarization around the charged groups.
\section{Model}
For a nonlocal medium, the relation between the electric field,
$\vct{E}$, and the displacement field, $\vct{D}$, is given by an
integral equation
$\vct{D}(\vct{x})=\int_\Omega \tu{d}\vct{x}'\bm{\epsilon}(\vct{x},
\vct{x}')\vct{E}(\vct{x}')$, where $\vct{x}$ and $\vct{x}'$ are two
position vectors, and $\bm{\epsilon}(\vct{x}, \vct{x}')$ represents a
tensor field which describes the nonlocal response of the dielectric
material over the domain $\Omega$. As a consequence, and in contrast
to a local medium, the nonlocal form of Gauss's law is now expressed
by an integrodifferential equation.  In general, the degree
of difficulty of solving such a problem largely depends on the form of
the dielectric function. It has been shown that a Lorentzian
approximation \cite{kornyshevChemSol1985} for the wave-vector
dependence of $\bm{\epsilon}(\vct{x}, \vct{x}')$ in Fourier space,
  \begin{equation}
  \widetilde{\epsilon}(k) = \epsilon_\infty+
  \frac{\epsilon_\infty (\theta-1)}{(\xi \sqrt{\theta})^2}
  \left(\frac{1}{k^2+(\xi \sqrt{\theta})^{-2}}\right),
\end{equation}
permits transforming the integrodifferential equation into a system
of two coupled second order differential equations
\cite{hildebrandt2004}. Here, $k =\lvert \vct{k} \rvert$ represents
the magnitude of the wave-vector $\vct{k}$ in Fourier space. In this
work we employ a nonlinear mean-field theory of nonlocal
electrostatics that relies on a two-field description of electrostatic
potential and polarization degrees of freedom
\cite{paillusson2010}. The model introduces an additional length scale
characterizing polarization correlations in the medium and thus
provides a means to study the effect of surface polarization charges
which may exist independently of an electric field at an
interface. The view has proven effective in explaining a range of
phenomena which are not addressed within the standard PB theory. For
example, the reduction of dielectric constant near the interfaces
emerges as a natural consequence of this framework
\cite{sahin2014}. Recently, the theory has been further extended to
account for various properties of electrolytes near charged surfaces
such as finite ion size \cite{schaaf2015}, or the oscillatory behavior
of the polarization field near an interface
\cite{hedley2023}. Furthermore comparisons of atomistic models and
experiments have confirmed the utility of the model within the weak
coupling regime where ion-ion correlations are negligible and PB
theory applies \cite{schaaf2015,monet2021,hedley2023}.

\subsection{Two-Field Formulation of Nonlocal Electrostatics}
We begin with the nonlocal formulation given by
Ref. \onlinecite{hildebrandt2004,paillusson2010}. In this formulation,
the electric field, $\vct{E}$ and the displacement field, $\vct{D}$
are described by the gradients of potential fields $\phi$ and $\psi$,
respectively; i.e. $\vct{E} = -\bm{\nabla} \phi$ and
$\vct{D} = -\bm{\nabla} \psi$.  The governing equations are
\begin{equation}
  \label{eq:psi-dim}
-\nabla^2 \psi = \rho,
\end{equation}
\begin{equation}
    \label{eq:phi-dim}
  -\epsilon_0 \epsilon_\infty \nabla^2 \phi +
  \frac{\epsilon_0 \epsilon_\tu{s}}{\lambda^2} \phi - \rho =
  \frac{\psi}{\lambda^2},
\end{equation}
where, $\lambda$ represents the polarization screening length in the
absence of ions, and $\rho = \rho_\tu{ion}+\varrho$ denotes the
total charge density due to ions and fixed charges. Furthermore,
considering a $1:1$ electrolyte, we assume that the ionic charge
density is given by the Boltzmann distribution,
\begin{equation}
  \rho_{\tu{ion}} = -2 n_0 e \sinh \left(
    \frac{e \phi}{k_\tu{B} T}
  \right).
\end{equation}
It is worth noting that the above formulation implicitly assumes that
the total polarization $\vct{P}_\tu{t}$ consists of two parts, namely
the configurational polarization, $\vct{P}$, and the atomic
polarization, $\vct{P}_\tu{a}$. The latter may be characterized by a
local response given by
$\vct{P}_\tu{a} = \epsilon_0 \chi_\infty \vct{E}$, where
$\chi_\infty = \epsilon_\infty -1$ is the susceptibility of the medium
arising from, e.g., electronic modes. Consequently, we have 
\begin{equation}
  \label{eq:disp}
  \vct{D} = \epsilon_0 \vct{E} + \vct{P}_\tu{t} =
  \epsilon_0 \epsilon_\infty \vct{E} + \vct{P},
\end{equation}
which permits the derivation of boundary conditions for $\phi$ and
$\psi$. 

For a charge-carrying surface in contact with an electrolyte,
the displacement field satisfies $\vct{D} \cdot \vct{n} = -\sigma$,
where $\sigma$ is the surface charge density and $\vct{n}$ is the
outward normal vector as shown in Fig. \ref{fig:schm}. This equation
provides a Neumann boundary condition for $\psi$;
i.e. $\bm{\nabla}\psi \cdot \vct{n} = \sigma$. Assuming a fixed
polarization surface charge, $\sigma_\tu{p} = \vct{P} \cdot \vct{n}$,
and using Eq. (\ref{eq:disp}), we derive another Neumann condition for
$\phi$, such that
$\bm{\nabla}\phi \cdot \vct{n} =
(\sigma+\sigma_\tu{p})/\epsilon_0\epsilon_\infty$.
\subsection{Dimensionless Governing Equations}
In the subsequent analysis we present the equations in the more
convenient dimensionless form. We scale distance by the Debye
screening length $\kappa^{-1}$.  Accordingly,
$\vct{x}_* = \kappa \vct{x}$ and $\bm{\nabla}_* = \kappa \nabla$
denote the scaled position vector and dimensionless gradient operator,
respectively. Scaling the potentials $\phi$ and $\psi$ by
$\phi_\tu{0} = k_\tu{B} T/e$ and
$\psi_0 = \epsilon_0 \epsilon_\tu{s} \phi_0$, respectively, the
governing equations can be written as
\begin{equation}
  \label{eq:psi-nondim}
  -\nabla_*^2 \psi_* = \rho_*,
\end{equation}
\begin{equation}
  \label{eq:phi-nondim}
  -\delta^2 \nabla_*^2 \phi_* =
  \delta ^2 \theta \rho_* + \theta (\psi_*-\phi_*),
\end{equation}
where $\delta = \kappa \lambda$ is the ratio of the polarization
screening length, $\lambda$, to the Debye screening length,
$\kappa^{-1}$, and the parameter
$\theta = \epsilon_\tu{s}/\epsilon_{\infty}$ denotes the ratio of the
dielectric constants at low and high frequencies. Similarly, the
rescaled charge density $\rho_*$ is given by
\begin{equation}
  \rho_* = -\sinh\phi_* + \varrho_*,
\end{equation}
where $\varrho_*$ is the dimensionless charge density due to the fixed
charges, e.g., embedded interfacial dipoles. The boundary conditions
in dimensionless form are thus (i)
$\bm{\nabla}_* \psi_* \cdot \vct{n}= \Sc$ and (ii)
$\bm{\nabla}_* \phi_* \cdot \vct{n} = \theta (\Sc+\Pc)$, where
$\Sc = 2 \sgn(\sigma)/ \kappa \ell_{c}$ and
$\Pc = \sgn(\sigma_\tu{p})/ \kappa \ell_{p}$ are dimensionless
parameters characterizing the surface charge densities $\sigma$ and
$\sigma_\tu{p}$, respectively. Here,
$\ell_\tu{c} = e/2\pi|\sigma|\ell_\tu{B}$ and
$\ell_{p} = e/2\pi|\sigma_\tu{p}|\ell_\tu{B}$ are Gouy-Chapman lengths
associated with $\sigma$ and $\sigma_\tu{p}$, respectively. 

The above rescaling demonstrates that the problem may be cast in the
form of an interplay between several characteristic length
scales. Importantly, the parameter $\delta$ captures the degree of
nonlocality in the medium. In the limit $\delta \rightarrow 0$, the
solution to Eqs.  (\ref{eq:psi-nondim})-(\ref{eq:phi-nondim})
converges to that of the standard PB equation that describes
electrostatics in a local medium. In the subsequent sections, we only
work with the dimensionless formulation, and henceforth, we drop all
$*$ symbols for simplicity.

\section{Problem Setup}
\begin{figure}[t!]
  \includegraphics[width=0.45\textwidth,
  trim={0.0cm -0.5cm 0cm 0cm},clip]{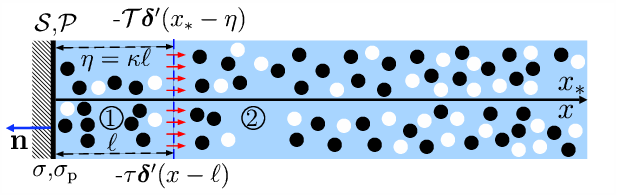}
  \caption{Schematic representation of an infinite flat plate carrying
    electrical charge and polarization charge densities given by
    $\sigma$ and $\sigma_\tu{p}$ respectively, in contact with a
    semi-infinite nonlocal electrolyte containing cations and
    anions. A negative value of $\sigma_\tu{p}$ corresponds to a
    polarization $P$ in the positive $x$-direction. A layer of dipoles
    of density $\tau$ may be embedded at $x = \ell$ (red arrows).}
  \label{fig:schm}
\end{figure}

We now consider a semi-infinite one-dimensional electrolyte in contact
with a charged surface at $x = 0$ (Fig. \ref{fig:schm}). In the
presence of a dipole layer at a distance $\eta = \kappa \ell$ from the
surface, the charge density, $\varrho$, is expressed by the derivative
of a one-dimensional $\delta$-function; i.e.
$\varrho = - \Tc \bm{\delta}'(x-\eta)$. Here,
$\Tc = 4 \pi \tau \ell_\tu{B}/e$ represents the dimensionless surface
dipole density of the dipole layer. The pertinent governing equations
are
\begin{equation}
  \label{eq:psi}
  - \frac{\tu{d}^2 \psi}{\tu{d} x^2} = \rho,
\end{equation}
\begin{equation}
  \label{eq:phi}
  \delta^2 \frac{\tu{d}^2 \phi}{\tu{d} x^2} = 
  \theta (\phi-\psi) -\theta \delta^2 \rho,
\end{equation}
where $\rho = -\sinh \phi - \Tc \bm{\delta}'(x-\eta)$ is the total
charge density due to ions and the dipole layer, and $\bm{\delta}(x)$
represents the Dirac delta function which is distinct from the
nonlocality parameter $\delta$. The boundary conditions for the
semi-axis, $x \in [0, \infty)$, are (i) $\psi'(0)=- \Sc$, (ii)
$\phi'(0)=- \theta(\Sc+\Pc)$, (iii) $\psi'(x) \rightarrow 0$ as
$x \rightarrow \infty$, and (iv) $\phi'(x) \rightarrow 0$ as
$x \rightarrow \infty$.

\section{Results and Discussion}
We explore the role played by interfacial polarization, and dipoles
embedded near the object-electrolyte interface, in the nonlocal
electrostatics problem by considering two cases: (1) $\Pc \neq 0$;
$\Tc = 0$ and (2) $\Pc = 0$; $\Tc \neq 0$, respectively. For small
electrical potentials, $\phi \ll 1$, the solution of Eqs.
(\ref{eq:psi}) and (\ref{eq:phi}) can be expressed in closed form
\cite{paillusson2010}.  Using the Debye-H\"uckel approximation,
$\rho \approx -\phi - \Tc \bm{\delta}'(x-\eta)$, we find that the
electrical potential, $\phi(x)$, has a double-exponential form
\begin{equation}
  \label{eq:DHsol}
  \phi(x) = A_+ \exp(-\kappa_+ x) + A_- \exp(-\kappa_- x),
\end{equation}
where
\begin{equation}
  \label{eq:kappa}
  \kappa_\pm = \left[\frac{\theta (1+\delta^2)}{2\delta^2}
    \left( 1 \pm
      \sqrt{1-\frac{4\delta^2}{\theta (\delta^2+1)^2}} \right)
    \right]^{1/2}
\end{equation}
represent dimensionless forms of two effective screening lengths
$\kappa_{1,2}=\kappa\kappa_\pm$ that arise on account of the coupling
between polarization charges and ions.  In the absence of interfacial
dipoles ($\Tc=0$), Eq. (\ref{eq:DHsol}) is valid over the entire
domain $x \ge 0$. In the presence of the dipole layer however, the
above equation is valid only in the region $x \ge \eta$. Accordingly,
the coefficients $A_+$ and $A_-$ are given by
\begin{equation}
  \label{eq:coef}
  A_i = \omega_i^{\tu{S}} \Sc + \omega_i^{\tu{T}} \Tc +
\omega_i^{\tu{P}} \Pc,
\end{equation}
where the functions
\begin{equation}
  \label{eq:omegaP}
  \omega_i^{\tu{P}} =
  \frac{\kappa_i\theta}{\kappa_i^2-\kappa_j^2},  
\end{equation}
\begin{equation}
  \label{eq:omegaS}
  \omega_i^{\tu{S}} = \frac{ \kappa_i
    (\theta-\kappa_j^2)}{\kappa_i^2-\kappa_j^2},
\end{equation}
and
\begin{equation}
  \label{eq:omegaT}
  \omega_i^{\tu{T}} =
  \frac{\kappa_i^2 \sinh (\kappa_i \eta) (\theta-\kappa_j^2)}
  {\kappa_i^2-\kappa_j^2},
\end{equation}
determine the contributions of $\Pc$, $\Sc$ and $\Tc$ to $A_i$,
respectively. In the equations above, we have $i,j \in \{+,-\}$ and
$i \neq j$. For $\theta \gg 1$, which may often be the case for polar
solvents, $\kappa_\pm$ may be well approximated by the expressions
$\kappa_+^2 \sim \theta (\delta^2 +1)/\delta^2$ and
$\kappa_-^2 \sim 1/(\delta^2+1)$, respectively.

\begin{figure}[ht!]
  \begin{center}
    \includegraphics[height=0.38\textwidth,
    trim={0.4cm  0.5cm 0.05cm 0.4cm},clip]{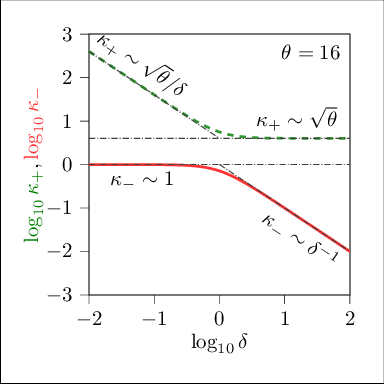}
  \end{center}
  \caption{
    Dependence of the two screening lengths, $\kappa_\pm$, on the
    nonlocality parameter, $\delta$, calculated for $\theta = 16$.}
  \label{fig:kappa}
\end{figure}

Examination of the above
asymptotic approximations shows that for a weakly nonlocal
medium ($\delta \ll 1$), $\kappa_+ \sim \sqrt{\theta}/\delta$ and
$\kappa_- \sim 1$, while for a strongly nonlocal medium
($\delta \gg 1$), we find that $\kappa_+ \sim \sqrt{\theta}$ and
$\kappa_- \sim 1/\delta$ (see Fig. \ref{fig:kappa}).
Since $\kappa_+>\kappa_-$, the coefficient $A_-$ determines the
electrical potential at long range. In particular, a distant observer
may interpret $A_-$ as the surface potential $\phi_\tu{s}$
characterizing the object whose far-field potential may be described
by a single exponential of the form
$\phi(x)=\phi_{\tu{s}}\exp(-\kappa_- x)$.  In the nonlinear regime,
where Eqs. (\ref{eq:psi}) and (\ref{eq:phi}) can only be solved
numerically, we investigate the behavior of different systems by
examining values of the effective surface potential $\phi_{\tu{s}}$
determined by fitting an exponential function
$\phi(x)=\phi_{\tu{s}}\exp(-\kappa_- x)$ to the far-field of
numerically calculated potentials $\phi(x)$ (see the supplementary
material). In the DH regime, we have $\phi_\tu{s} = A_-$ which is
given by Eq. (\ref{eq:coef}). It is worth noting here that the
polarization and dipole contributions to $A_-$, as given by
Eqs. (\ref{eq:coef})-(\ref{eq:omegaT}), immediately suggest the
possibility of a departure in sign of $\phi_\tu{s}$ from that given by
electrical charge alone.

\begin{figure}[ht!]
  \begin{center}
    \includegraphics[width=0.4\textwidth,
    trim={0.0cm  0.cm 0.cm 0.cm},clip]{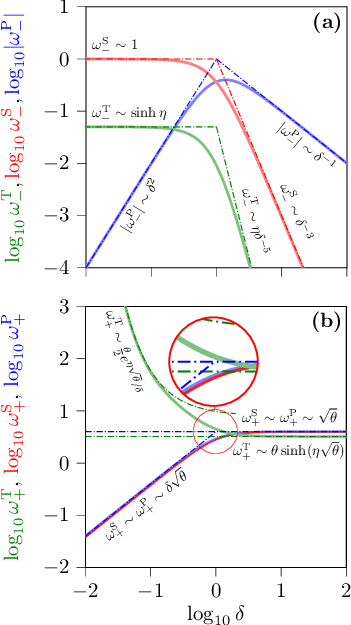}
  \end{center}
  \caption{Dependence of the functions $\omega_i^j$, given
      by Eqs. (\ref{eq:omegaP})-(\ref{eq:omegaT}), on the nonlocality
      parameter, $\delta$, calculated for $\theta = 16$ and
      $\eta = 0.05$. (a) shows the behavior of the long-range
      coefficients $\omega_-^{\mathrm{S}}$, $\omega_-^{\mathrm{P}}$,
      and $\omega_-^{\mathrm{T}}$ in different nonlocality regimes.
      (b) displays the dependence of the short-range coefficients
      $\omega_+^{\mathrm{S}}$, $\omega_+^{\mathrm{P}}$, and
      $\omega_+^{\mathrm{T}}$ on $\delta$.}
  \label{fig:omega}
\end{figure}

For a more comprehensive understanding of the overall response, we
analyze the behavior of the coefficients $A_-$ and $A_+$ in different
regimes of nonlocality using Eqs. (\ref{eq:coef})-(\ref{eq:omegaT}).
For weakly nonlocal media ($\delta \ll 1$), also called the local
limit, the functions $\omega_-^{\tu{S}} \sim 1$, and
$\omega_-^{\tu{T}} \sim \sinh \eta$ dominate the contribution to the
far-field potential while the function
$\omega_-^{\tu{P}} \sim -\delta^2$ vanishes as $\delta \rightarrow 0$
(see Fig. \ref{fig:omega}a). Furthermore, in the local limit, the
far-field potential is entirely determined by the electrical charge if
the distance between the dipole layer and the interface, $\ell$, is
significantly smaller than the Debye screening length, $\kappa^{-1}$;
i.e.  $\eta = \kappa \ell \ll 1$. A similar analysis reveals that
$\omega_+^{\tu{S}} \sim \omega_+^{\tu{P}} \sim \delta \sqrt{\theta}$
and $\omega_+^{\tu{T}} \sim \theta \exp(\eta \sqrt{\theta}/\delta)/2$
as $\delta \rightarrow 0$ (see Fig. \ref{fig:omega}b). From
Eq. (\ref{eq:DHsol}) and the fact that
$\kappa_+ \sim \sqrt{\theta}/\delta$ in this regime, we notice that
$A_+ \exp(-\kappa_+ x)$ vanishes for all $x>\eta$, as
$\delta \rightarrow 0$. In other words, the electrical potential,
$\phi$, recovers the PB solution in the DH regime as expected, i.e.,
$\phi(x) \rightarrow (\Sc+\Tc \sinh \eta) \exp(-x)$.

On the other hand, for a strongly nonlocal medium, $\delta \gg 1$, we
find that $\omega_-^{\tu{P}} \sim -\delta^{-1}$,
$\omega_-^{\tu{S}} \sim \delta^{-3}$, and
$\omega_-^{\tu{T}} \sim \eta \delta^{-5}$ (see
Fig. \ref{fig:omega}a). This implies that all contributions due to
$\Pc$, $\Sc$, and $\Tc$ are vanishingly small as
$\delta \rightarrow \infty$.  In fact as $\delta \rightarrow \infty$,
the long-range term $A_- \exp(-\kappa_- x)$ vanishes, and the
short-range term $A_+ \exp(-\kappa_+ x)$ converges to a linear PB
solution for a surface carrying an effective charge
$\sigma_\tu{eff}=\sigma+\sigma_\tu{p}$, and a medium of uniform
dielectric constant $\epsilon_\infty$. This can be verified by
analysis of the short-range coefficients, which shows
$\omega_+^{\tu{S}} \sim \omega_+^{\tu{P}} \sim \sqrt{\theta}$, and
$\omega_+^{\tu{T}} \sim \theta \sinh(\eta \sqrt{\theta})$ when
$\delta \gg 1$ (Fig. \ref{fig:omega}b). We note however,
that for practically relevant values $\delta >1$, the contribution
from the polarization surface charge, $\Pc$, dominates the far-field
potential.

\begin{figure}[t]
  \begin{center}
    \includegraphics[height=0.22\textwidth,
    trim={0.1cm 1.5cm 2.4cm 0.1cm},clip]{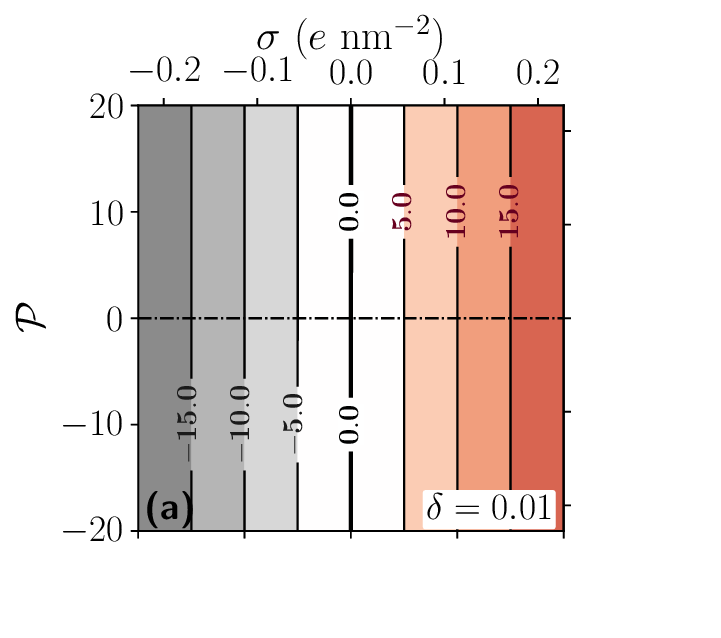}
    \includegraphics[height=0.22\textwidth,
    trim={1.5cm  1.5cm 0.5cm 0.1cm},clip]{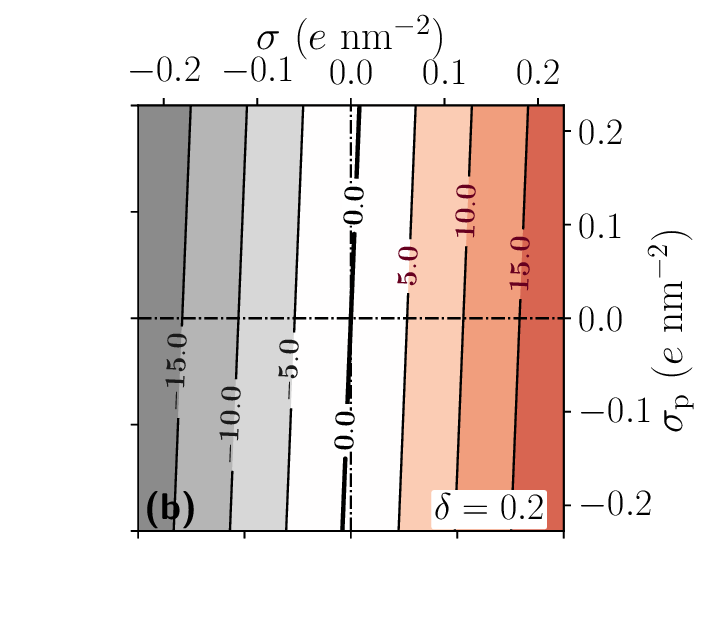} \\
    \includegraphics[height=0.22\textwidth,
    trim={0.1cm 0.1cm   2.4cm 1.5cm},clip]{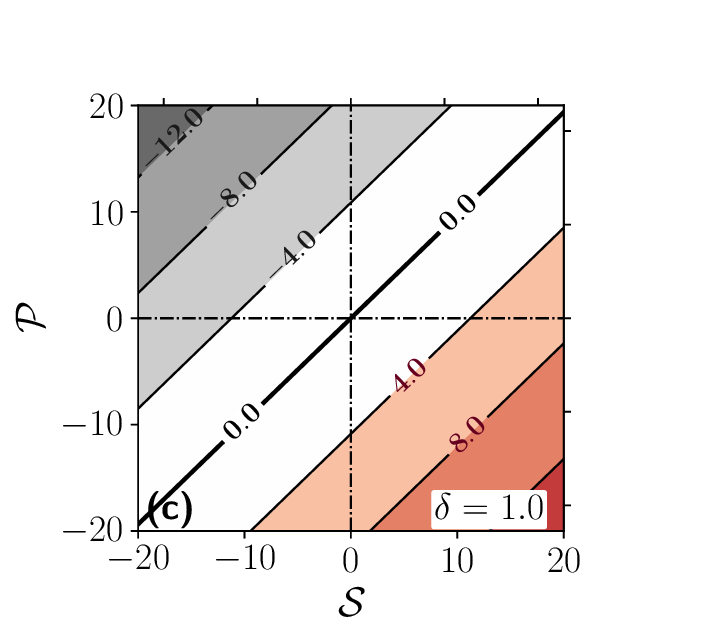}
    \includegraphics[height=0.22\textwidth,
    trim={1.5cm  0.1cm 0.5cm 1.5cm},clip]{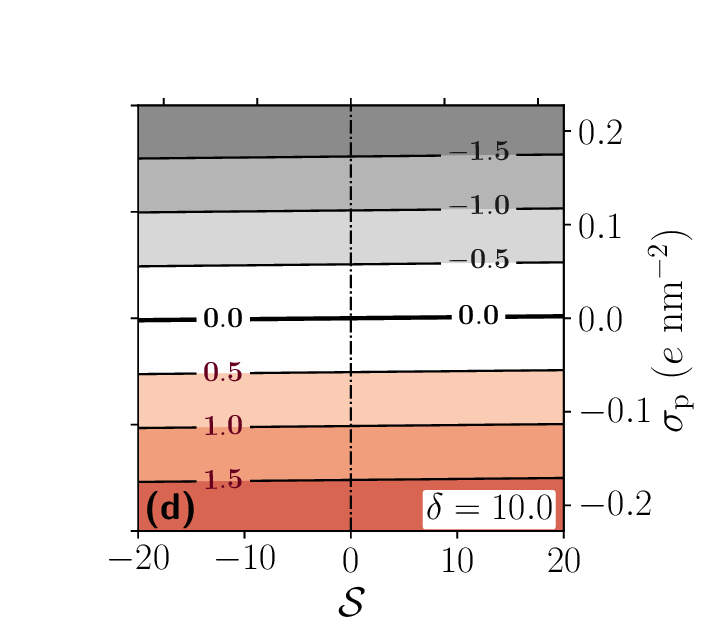}
  \end{center}
  \caption{Contours of effective surface potential,
    $\phi_\tu{s} = A_-$, for the $(\Sc, \Pc)$ problem in DH regime
    calculated using a Bjerrum length of $\ell_\tu{B} \approx 0.7$ nm,
    characteristic of water at $298$ K, and Debye screening length
    $\kappa^{-1} = 10$ nm. }
  \label{fig:PDH}
\end{figure}

In general the dependence of $\phi_\tu{s}$ on a pair of parameters
$(\Sc,\Pc)$ or $(\Sc,\Tc)$ may be visualized as two-dimensional
contour plots which reveal interesting trends.  First, for
$\Pc \neq 0$; $\Tc = 0$ (dipole layer absent), we observe that in
the DH regime, the contourlines of $A_-$ are described by parallel
straight lines and the zero-level contour, given by
$\Pc=(\kappa_+^2/\theta-1)\Sc$, divides the $\Sc$--$\Pc$ plane into
two regions of positive and negative effective surface potentials
(Fig. \ref{fig:PDH}). When $\delta \rightarrow 0$, this line coincides
with the $\Pc$-axis capturing the independence of $A_-$ from $\Pc$
(Fig. \ref{fig:PDH}a).
Thus in a local medium, the sole contribution to the far-field
electrical potential stems from electrical charge. Such a behavior can
be explained by the fact that in the limit $\lambda \to 0$ the rapid
spatial variation of polarization, $\vct{P}$, gives rise to a
volumetric polarization charge,
$\rho_\tu{p} = - \bm{\nabla} \cdot \vct{P}$, which effectively screens
the polarization surface charge $\Pc$ over short distances thereby
diminishing its contribution to the far field potential. As $\delta$
grows in magnitude however, the zero-level contour rotates clockwise
about the origin, and the interfacial polarization begins to
contribute significantly to the far-field electrical potential. At
intermediate values of $\delta$ there exist regions of the parameter
space where the surface polarization, $\Pc$, can compete with surface
charge, $\Sc$, and even invert the sign of the effective surface
potential $\phi_\tu{s}$ (Fig. \ref{fig:PDH}).  For $\delta \gg 1$,
$\phi_\tu{s}$ becomes independent of $\Sc$, and the behavior of the
system in the far-field is purely determined by the polarization
surface charge density, $\Pc$, which is diametrically opposite to
behavior encountered in a local medium.  While the above analysis
provides qualitative insights into the impact of nonlocal effects on
electrical potentials in electrolytes, the magnitude of the effective
surface potential, $A_-$, in Fig. \ref{fig:PDH} implies that in
general the problem lies significantly outside the regime of the
validity of the DH approximation, $\phi \ll 1$. We therefore examine
the problem in the nonlinear regime by numerically solving Eqs.
(\ref{eq:psi})-(\ref{eq:phi}).

\begin{figure}[t!]
 \begin{center}
   \includegraphics[height=0.22\textwidth,
   trim={0.1cm 1.5cm 2.4cm 0.1cm},clip]{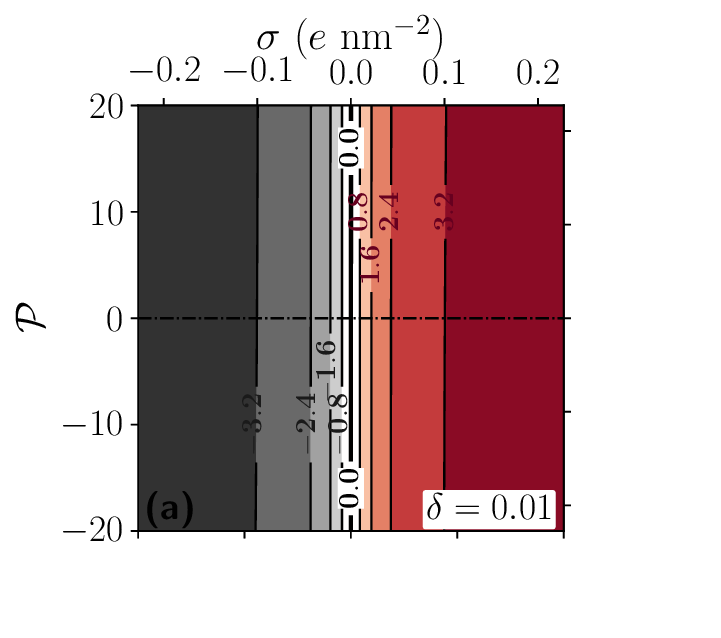}
   \includegraphics[height=0.22\textwidth,
   trim={1.5cm  1.5cm 0.5cm 0.1cm},clip]{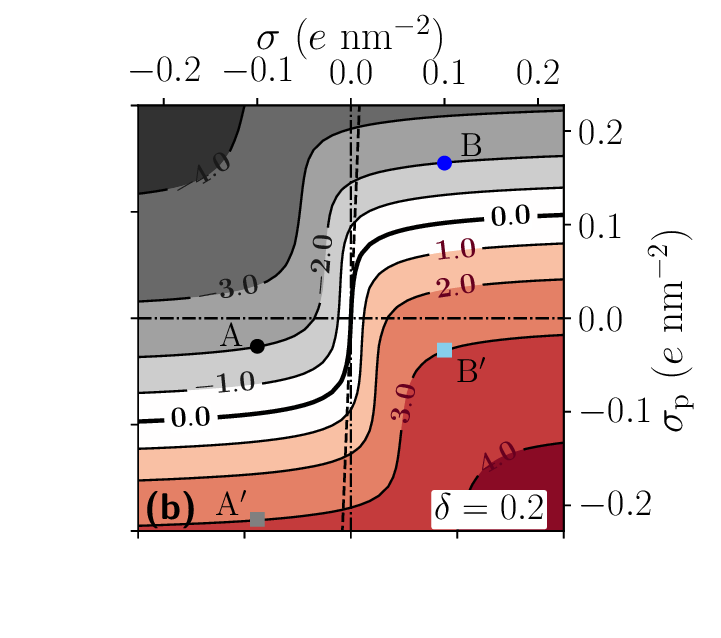} \\
   \includegraphics[height=0.22\textwidth,
   trim={0.1cm 0.1cm   2.4cm 1.5cm},clip]{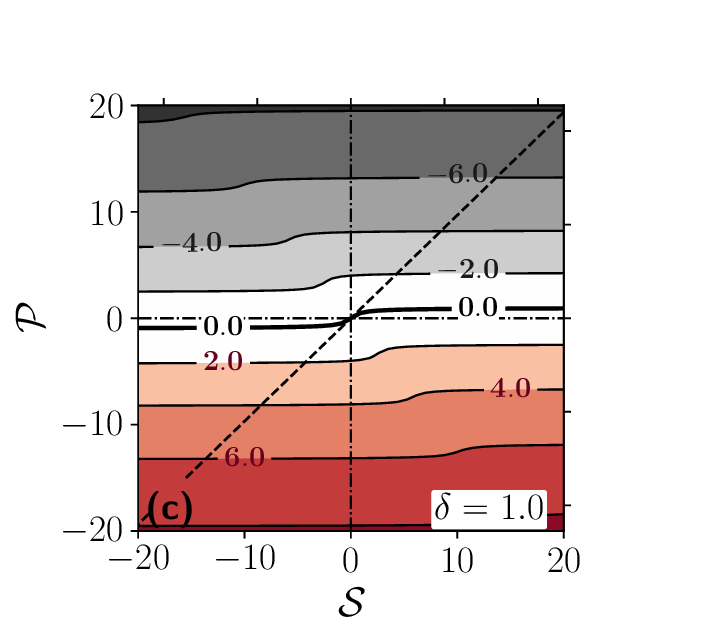}
   \includegraphics[height=0.22\textwidth,
   trim={1.5cm  0.1cm 0.5cm 1.5cm},clip]{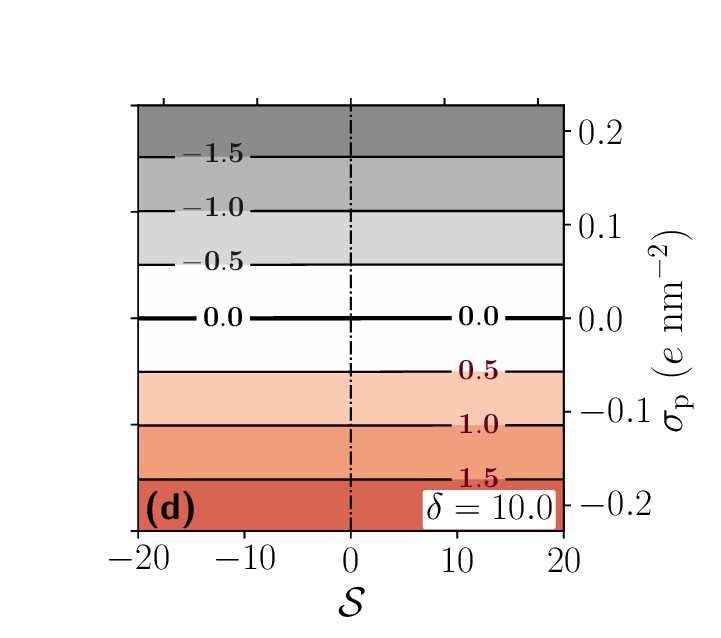}
   \end{center}
   \caption{Contours of effective surface potential, $\phi_\tu{s}$,
     for the $(\Sc, \Pc)$ problem in the nonlinear regime, calculated
     using a Bjerrum length of $\ell_\tu{B} \approx 0.7$ nm,
     characteristic of water at $298$ K, and Debye screening length
     $\kappa^{-1} = 10$ nm. Dashed lines in (b) and (c) depict the
     zero-level contour in the DH regime.}
  \label{fig:PNL}
\end{figure}

The results of the analysis in the nonlinear regime reveal trends that
are qualitatively comparable to the linear regime but are more
multifaceted in general. In the local limit ($\delta \rightarrow 0$),
the non-uniform density of contourlines of fixed step size reflects
the emergence of nonlinearity as $\Sc$ increases
(Fig. \ref{fig:PNL}a).  As $\delta$ grows, a substantial change in the
shape of zero-level contour leads to a significant change in the
far-field behavior. For $\delta \geq 0.2$ and a wide range of $\Sc$
values, the sign of the far-field electrical potential depends
sensitively on the value of $\Pc$ (Fig. \ref{fig:PNL}). In fact, for
reasonable values of both $\Sc$ and $\Pc$ ($\lvert\sigma\rvert$ and
$\lvert\sigma_\tu{p}\rvert\approx 0.1 e\tu{/nm^2}$) the sign of the
potential can be {\it opposite} to that expected based on the
electrical charge alone.

\begin{figure}[t!]
  \begin{center}
    \includegraphics[height=0.75\textwidth,
    trim={0.0cm  0.0cm 0.00cm 0.0cm},clip]{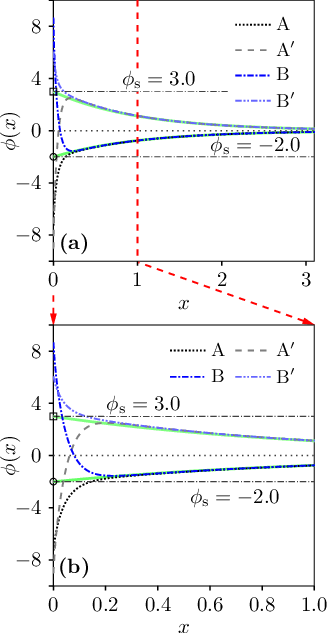}    
  \end{center}
  \caption{Profiles of electrical potential. (a) Electrical potential
    $\phi(x)$, for points $\tu{A}$, $\tu{A}'$, $\tu{B}$, and $\tu{B}'$
    indicated in Fig. \ref{fig:PNL}b. Solid green lines represent fits
    to the far-field potential profile
    $\phi(x)=\phi_{\tu{s}}\exp(-\kappa_-x)$ where
    $\kappa\kappa_-=\kappa_{\tu{eff}}$ represents the effective
    screening length. (b) Magnification of the short-range behavior of
    $\phi(x)$ signifying the disparities between the systems $\tu{A}$,
    $\tu{A}'$, $\tu{B}$, and $\tu{B}'$ near the surface.}
    \label{fig:potDist}
  \end{figure}

To emphasize this point, we consider systems on either side of the
zero-contour line in a pairwise fashion, e.g., characterized by points
$\tu{A}$, $\tu{A}'$, $\tu{B}$ and $\tu{B}'$ in the parameter space
(Fig. \ref{fig:PNL}b). First we consider pairs of systems
$(\tu{A},\tu{A}')$ and $(\tu{B}, \tu{B}')$ such that within each pair,
$\Sc$ is identical both in sign and magnitude; i.e.
$\Sc_\tu{A}= \Sc_\tu{A'}$ and $\Sc_\tu{B}= \Sc_\tu{B'}$.  Furthermore
the magnitudes of the effective surface potential,
$\lvert\phi_\tu{s}\rvert\approx 2$--$3$, are also comparable. However,
importantly, the signs of $\phi_{\tu{s}}$ are opposite in each
pairing. Figure \ref{fig:potDist}a displays the spatial variation of
$\phi(x)$ together with single exponential fits used to determine
$\phi_{\tu{s}}$ in each case.  Similarly, pairs of systems given by
points $(\tu{A}',\tu{B}')$ and $(\tu{A},\tu{B})$ are characterized by
opposite signs of charge density, $\Sc$, and therefore display
different $\phi(x)$ profiles at short-range
(Fig. \ref{fig:potDist}b). But within each pair, the spatial
electrical potential profiles are essentially indistinguishable in the
far field. Taken together, the above observations imply that
electrically like-charged objects may appear oppositely charged in the
far-field, and equally, that oppositely charged entities may appear
like-charged at large separations (Fig. \ref{fig:potDist}a).

Finally, we examine the far-field surface electrical potential,
$\phi_{\tu{s}}$, in case (2), $\Tc\neq0$, $\Pc=0$, in the
($\Sc$--$\Tc$) plane for a dipole layer embedded near an interface in
the electrolyte.  This problem has been previously considered in the
linear regime and the contribution of the dipoles to the far field was
found to be negligible \cite{belaya1994a, belaya1994b}. This is in
sharp contrast to our results on a system described by a
\textit{polarization} boundary condition, described above.

Equation (\ref{eq:coef}) reveals a marked distinction between the
$(\Sc,\Pc)$ and $(\Sc,\Tc)$ problems in the DH regime. Mainly we find
that, keeping all other parameters constant, the contribution of a
dipole layer to the effective surface potential vanishes entirely as
$\eta \rightarrow 0$. This behavior, common to both local and nonlocal
media \cite{belaya1994a}, implies that the influence of a dipole layer
can only propagate into the far-field when it is embedded in the
electrolyte rather than located at the surface ($\ell=0$).
  
\begin{figure}[t!]
  \begin{center}
    \includegraphics[height=0.223\textwidth,
    trim={0.1cm 1.5cm 2.4cm 0.1cm},clip]{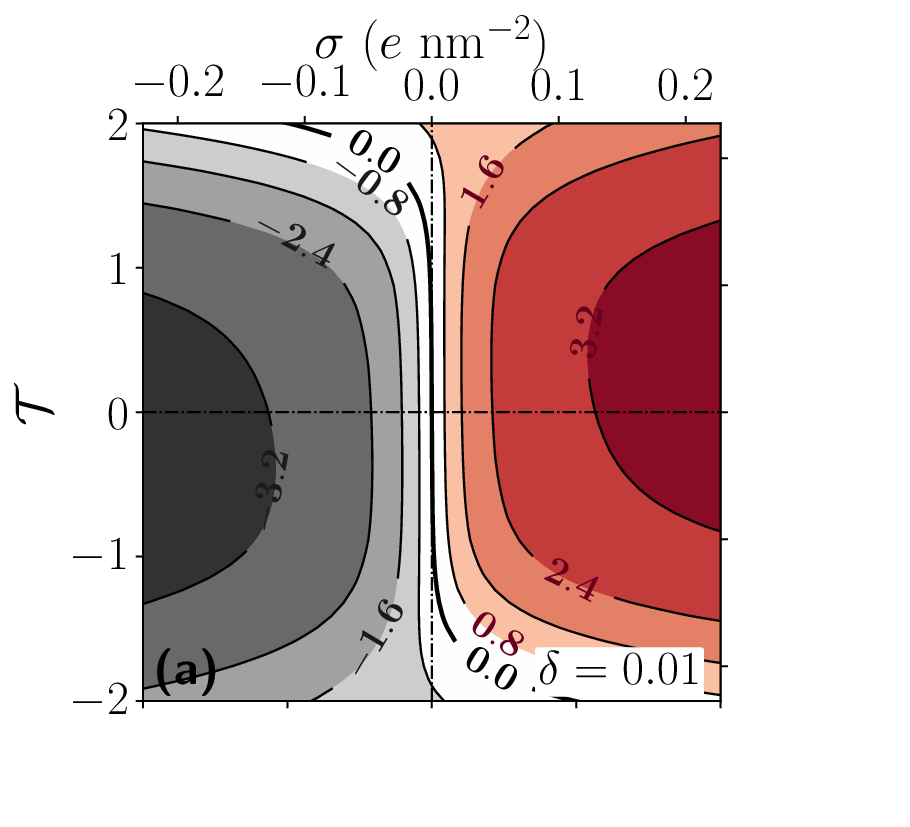}
    \includegraphics[height=0.223\textwidth,
    trim={1.5cm  1.5cm 0.5cm 0.1cm},clip]{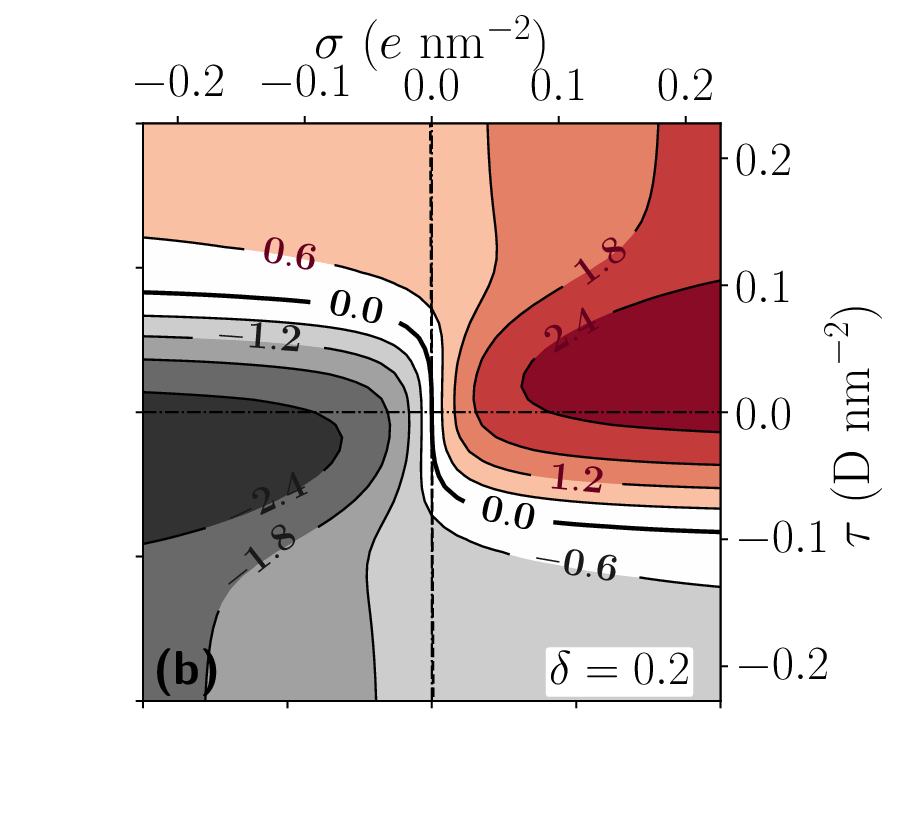} \\
    \includegraphics[height=0.223\textwidth,
    trim={0.1cm 0.1cm   2.4cm 1.5cm},clip]{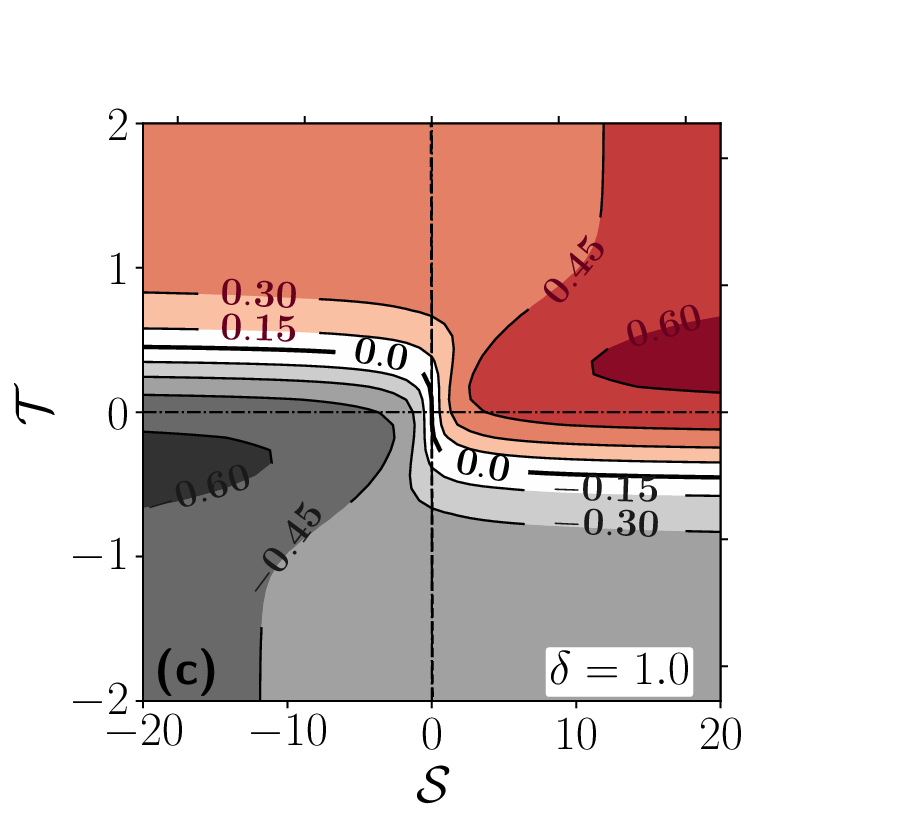}
    \includegraphics[height=0.223\textwidth,
    trim={1.5cm  0.1cm 0.5cm 1.5cm},clip]{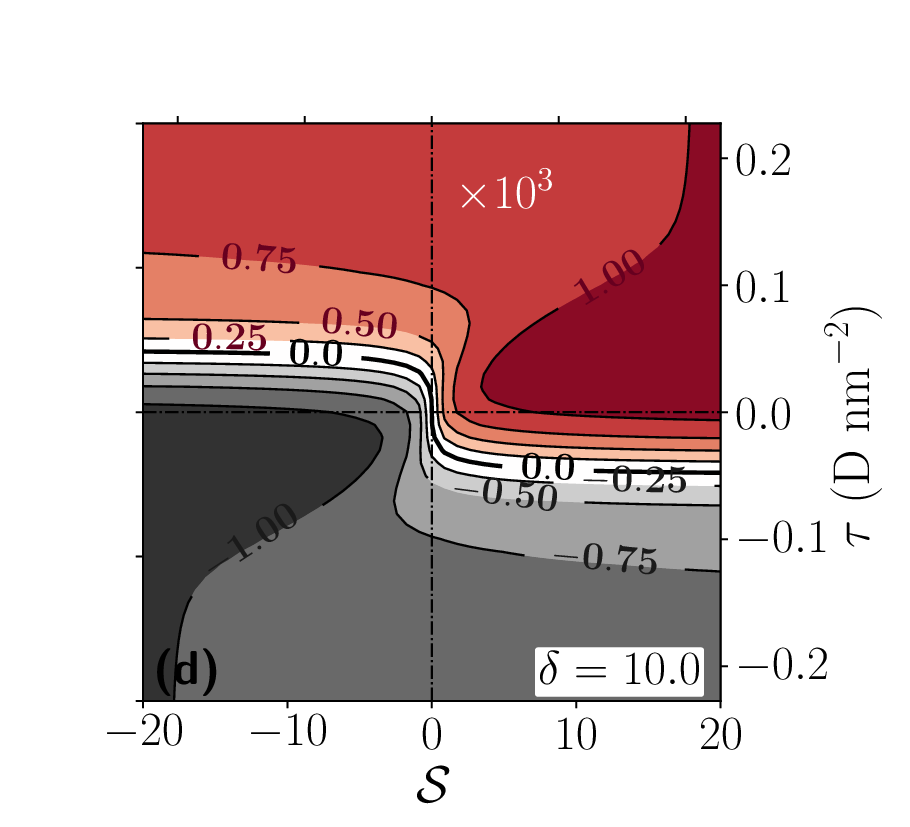}
   \end{center}
   \caption{Contours of effective surface potential, $\phi_\tu{s}$,
     for the $(\Sc, \Tc)$ problem in nonlinear regime calculated using
     a Bjerrum length of $\ell_\tu{B} \approx 0.7$ nm, characteristic
     of water at $298$ K, and Debye screening length
     $\kappa^{-1} = 10$ nm. The dipole layer is located at
     $\eta = 0.05$. Contour level values of $\phi_\tu{s}$ in (d) have
     been multiplied by a factor of $10^3$.}
     \label{fig:Tneq0} 
\end{figure}

Investigating the problem in the nonlinear regime for $\eta = 0.05$
(corresponding to $\ell = 0.5$ nm and $\kappa^{-1}=10$ nm), we find
that an interfacial dipole layer can indeed make a tangible
contribution to the far field for $\delta \gtrsim 0.2$, similar to the
situation with a polarization boundary condition (see
Fig. \ref{fig:Tneq0}).  But the trends observed for interfacial
dipoles do reflect important quantitative differences to the
polarization case that are most significant in the regime of
$\delta \gg 1$. Here, surprisingly, neither the electrical charge nor
electrical dipoles contribute to the electrical potential which nearly
vanishes in the far-field. This result may be best understood by the
preceding asymptotic analysis in the linear regime which reveals that
$A_-$ decreases rapidly as $\delta\to\infty$.

\section{Conclusion}
A central finding of our work is the unexpected role of the
polarization boundary condition in influencing the sign of the electrical
potential in the far field, as illustrated in our discussion of
Figs. (\ref{fig:PNL}) and (\ref{fig:potDist}). This observation may
have important ramifications for the counterintuitive and apparently
anomalous experimental reports of ``like-charge attraction'' in the
literature over the past several decades.  Attraction
between non-identical but electrically like-charged objects may
emerge as a natural consequence of this theoretical view, for
experimentally meaningful values of all parameters. By the same token,
``opposite-charge repulsion'' may also be expected. These results also
support the possibility of experimentally observing non-zero
electrophoretic mobilities as well as far-field attraction or
repulsion for formally net charge-neutral objects and surfaces
\cite{egawa1999,silbert2012,chibowski2016}. 

A key result of nonlocal models
\cite{gruen1983a,gruen1983b,belaya1987,
  belaya1994b,rucken2002,paillusson2010,blossey2023} is the existence
of two inverse screening lengths, $\kappa_{1,2}$, where both depend on
the length scale $\lambda = \xi \sqrt{\theta}$ and
$\sqrt{\theta} \approx 4$--$9$ in the present approach. Given the
significant influence of $\lambda$ on both the qualitative behavior of
the electrical potential as well as the magnitude of the overall
effective screening length in the far-field, an examination of the
microscopic sources and mechanisms that govern the value of $\xi$ is
central to an assessment of the importance of nonlocal effects in
practical contexts. Models of nonlocal electrostatics have
traditionally attributed a variety of underpinning physical sources to
the length scale $\xi$. These range from the most obvious: the length
scale of molecular rotational correlations given by the intermolecular
separation of $2$--$3$ \AA, to distances derived from interfacial
dipole layer separations ($\approx 1.5$ \AA) \cite{rucken2002}, or the
density of Bjerrum defects that act as sources and sinks of the
polarization field in the medium ($\approx 33$ nm for ice)
\cite{gruen1983a}. Based on a value of $\xi\approx 3$\AA~expected to
describe pure bulk water, $\lambda$ would not be expected to exceed
$2$--$3$ nm, which aligns reasonably well with literature estimates
\cite{rucken2002,rubashkin2014,hildebrandt2004}. This would in turn
imply the onset of nonlocality given by, e.g., $\delta >1$, and
symbolized by (1) the departure of $\kappa_\tu{eff}^{-1}$ from the
Debye screening length $\kappa^{-1}$, possibly accompanied by (2)
the inversion of the sign of the far-field electrical potential, at
concentrations as low as $10$ mM of monovalent salt
($\kappa^{-1} \approx 3$ nm).

Interestingly, nonlinear spectroscopy measurements of the decay length
scale of polarization fluctuations in aqueous media have reported
values as high as $\xi=20$ nm in water \cite{duboisset2018}, $3$--$5$
nm in electrolytes \cite{chen2016,duboisset2018,dedic2019}, and much
larger for polyelectrolytes in solution \cite{dedic2021}. The same is
true for a range of apolar media where hyper-Rayleigh light scattering
measurements suggest approximate length scales of $40$ nm
\cite{shelton2014} arising from phonon modes in the medium. The source
of these correlations in water -- whether they arise from poorly
understood properties of the hydrogen bonding network or simply from
the electric field due to ions in solution -- remains unclear
\cite{jungwirth2018}.  Regardless of their origin, polarization
correlation lengths of $\xi \approx 5$--$25$ nm in water would imply
polarization screening lengths, $\lambda\gtrsim50$ nm, which if true,
would entail significant repercussions in experiments. For instance at
salt concentrations higher than $0.1$ mM, where $\kappa^{-1}<30$ nm,
$\lambda\sim50$ nm entails $\delta\gtrsim1$, implying that
electrostatics can be significantly nonlocal under most experimentally
relevant conditions. This entails not only significantly longer
effective screening lengths but also possible inversion in the sign of
the measured far-field electrical potential.

The discussion on the origins and value of $\xi$ would not be complete
without insights from earlier reports, and specifically a recent MD
simulation study on the related problem of short-ranged hydration
forces in lipid bilayer systems
\cite{marcelja1976,kornyshev1989,leikin1990, schlaich2024}. Netz
\textit{et al.} have suggested that laterally inhomogeneous water
ordering can introduce a surface structural wavelength, which may
influence the overall range of these short-range forces. Under the
na\"ive assumption of translational invariance of surface properties
(a homogeneous surface), such length scales -- measured or inferred
from experiments or from molecular simulations -- would tend to be
attributed to the properties of bulk water. If the polarization
correlation and/or screening length is not solely governed by the
properties of the bulk fluid there may not be grounds to expect $\xi$
to be strictly limited to the rather small value given by the
intermolecular separation of bulk solvent molecules.  Although our
present framework is concerned solely with a laterally averaged,
homogeneous dipolar polarization density normal to the surface as the
relevant structural order parameter in a Landau-Ginzburg formulation,
higher order multipole densities, as well as their curls and
gradients, may also contribute to the electrical potential on similar
footing, each accompanied by an additional characteristic length
scale. Similarly, other microscopic features of the electrolyte such
as finite ion-size may be accounted for within the model as previously
described \cite{hedley2023}. Finally, given the extensive reports on
cooperative effects and ordering of water due to the presence of ions
and polyelectrolytes in electrolytes, it may even be conceivable to
entertain the idea of a length scale in the problem related to the
size of the defect embedded in the medium, be this an ion, a molecule
or a particle \cite{tielrooij2010,chen2016,dedic2019,dedic2021}. All
these considerations point to rather intricate possible microscopic
underpinnings of the relevant length scales $\xi$ and
$\lambda$. System-dependent and therefore non-universal governing
length scales suggest that nonlocal electrostatic effects may be
ubiquitous and more elaborate in character than previously
anticipated.

\section{Supplementary Material}
The details of the numerical simulations have been described in the
Supplementary Material.

\begin{acknowledgements}
  The authors gratefully acknowledge funding from the European
  Research Council (ERC) under the European Union’s Horizon 2020
  research and innovation programme (No 724180), and Roland Netz for
  helpful discussions.
\end{acknowledgements}

\section*{Author Declarations}
\noindent
\textbf{Conflict of Interest}

The authors have no conflicts to disclose.

\section*{data availability}
All data are available within the article or supplemental information.

\bibliography{ref.bib}

\end{document}